\documentclass[prd,twocolumn,showpacs,a4paper,superscriptaddress,nofootinbib]{revtex4}
\usepackage{graphicx,multirow,color,xspace,hyperref}
\usepackage{amsfonts,amsmath}
\usepackage{mathrsfs}
\usepackage[innercaption]{sidecap}

\newcommand{\dDx}{\text{d}^D\hspace{-0.4ex}x\,}

\begin{document}
\setcounter{topnumber}{1}
\title{The two faces of quantum sound}
\author{C. Barcel\'{o}}
\affiliation{Instituto de Astrof\'{\i}sica de Andaluc\'{\i}a (IAA-CSIC),
Glorieta de la Astronom\'{\i}a, 18008 Granada, Spain}
\author{L. J. Garay}
\affiliation{Departamento de F\'{\i}sica Te\'{o}rica II, Universidad Complutense
de Madrid, 28040 Madrid, Spain}
\affiliation{Instituto de Estructura de la Materia (IEM-CSIC), Serrano 121,
28006 Madrid, Spain}
\affiliation{King's College London, Department of Physics, Strand, London WC2R
2LS, UK}
\author{G. Jannes}
\affiliation{Instituto de Astrof\'{\i}sica de Andaluc\'{\i}a (IAA-CSIC),
Glorieta de la Astronom\'{\i}a, 18008 Granada, Spain}
\affiliation{Instituto de Estructura de la Materia (IEM-CSIC), Serrano 121,
28006 Madrid, Spain}
\affiliation{Universit\'{e} de Nice Sophia Antipolis, Laboratoire J.-A. Dieudonn\'{e}, UMR
CNRS-UNS 6621, Parc Valrose, 06108 Nice Cedex 02, France}

\date{\today}

\begin{abstract}
Fluctuations around a Bose-Einstein condensate can be described by
means of Bogolubov theory leading to the notion of quasiparticle and
antiquasiparticle familiar to non-relativistic condensed matter
practitioners. On the other hand, we already know that these
perturbations evolve according to a relativistic Klein-Gordon equation in
the long wavelength approximation. For shorter wavelengths, we show that this equation acquires nontrivial corrections which modify the Klein-Gordon product. In this approach, quasiparticles can
also be defined (up to the standard ambiguities due to observer-dependence). We demonstrate that---in the low as well as in the high energy
regimes---both concepts of quasiparticle are actually the same, regardless
of the formalism (Bogolubov or Klein-Gordon) used to describe
them. These results also apply to any barotropic, inviscid, irrotational
fluid, with or without quantum potential. Finally, we illustrate how the
quantization of these systems of quasiparticles proceeds by analyzing a
stationary configuration containing an acoustic horizon. We show that
there are several possible choices of a regular vacuum state, including a regular generalization
of the Boulware vacuum. Issues such us Hawking radiation crucially depend on
this vacuum choice.

\end{abstract}
\pacs{04.62.+v, 03.75.Kk, 04.70.Dy}
\maketitle

\section{Introduction}
It has for quite some time been understood that the propagation of sound
waves in inviscid irrotational barotropic fluids can, under rather general
circumstances,
 be effectively described by relativistic curved-spacetime metrics, see
e.g.~\cite{White:1973}. This led to the observation that it should, at least
in principle, be possible to achieve sonic or acoustic black hole
configurations, and thus study certain types of high-energy
effects on black hole physics by analogy~\cite{Unruh:1980cg}. This
gravitational analogy was established on firm foot about a decade ago
in~\cite{Visser:1997ux}. Since then, the field of analogue gravity has
grown into a mature and well-established research programme within the
gravity community~\cite{Barcelo:2005fc}. It offers both the exciting
prospect of bringing   black hole physics, in particular Hawking radiation,
within the reach of experimentation, as well as a refreshing conceptual
take on some long-standing problems of quantum gravity, such as its
relation to dark energy~\cite{Volovik:2004gi} or the avoidance of black
hole singularities in the gravitational collapse of ultra-heavy
bodies~\cite{Barcelo:2010vc}.

The quest for a useful background fluid in which to conduct experiments
of analogue gravity leads in the direction of superfluids.
Indeed, superfluids have a vanishing viscosity, and some superfluids can be made
extremely pure. Among the possible candidates,  Bose-Einstein
condensates~\cite{Garay:1999sk,Garay:2000jj}  have the comparative
advantage of being conceptually well understood, and relatively simple to
describe theoretically and manipulate experimentally. Acoustic black
holes  have  been recently reported for the first time in Bose-Einstein
condensates~\cite{Lahav:2009wx}, and there is good hope that an
experimental detection of analogue Hawking radiation, one of the main
current promises of the analogue gravity programme, will be achievable in the
near future~\cite{Carusotto:2008ep}.

Another motivation for studying Bose-Einstein condensates in the
context of analogue gravity is that they provide a real system in which
high-energy modifications to the relativistic dispersion relations arise.
This shows that there exist concrete examples in nature in which Lorentz
invariance is realized as a low-energy effective symmetry, broken at high
energy. Such a scenario could be the case for the local
Lorentz invariance of general relativity as well. In this manner,
Bose-Einstein condensates might offer an interesting model for quantum
gravity phenomenology, see e.g.~\cite{Liberati:2005id,Barcelo:2007ru}.

Our first aim  in this paper is to further elaborate and consolidate the
theoretical framework for the gravitational analogy in Bose-Einstein condensates by
taking a look at the different inner products and related creation and
annihilation variables that can be introduced depending on the point of
view. Quantum sound in Bose-Einstein condensates can, on the one hand,
be analyzed within the Bogolubov formalism by directly perturbing the
Gross-Pitaevskii equation. On the other hand, the phase
perturbations of the condensate obey a modified Klein-Gordon equation
and a corresponding quantization can be carried out. Remarkably, both
procedures give  rise to the same quantum theory. This allows us to establish a deep conceptual connection between both formalisms, the first one being inherently non-relativistic while the second is relativistic, up to corrections which are vanishingly small for long wavelengths. Both procedures are known to be equivalent in this long-wavelength acoustic limit, see e.g.~\cite{Kurita:2008fb}. Here we discuss in detail how the usual Klein-Gordon field is distorted after the acoustic approximation is broken. The step-by-step analysis that we perform shows that the equivalence between the Bogolubov approach and this generalized Klein-Gordon formalism persists well beyond the limit of validity of the acoustic approximation.

Once an inner product has been defined, one can proceed with the
quantization of the system of quasiparticles in the standard Fock manner.
An orthonormal and complete set of positive norm modes is needed
to expand the field operator. Finding such a complete set of positive norm
modes amounts to defining quasiparticle creation and annihilation operators
and a vacuum state. This construction is not unique: one can choose
several distinct sets of modes and vacuum states which give place to
different quasiparticle notions. To illustrate this procedure we will
consider a stationary one-dimensional configuration possessing an
acoustic black hole horizon. The second aim of this paper is precisely to show that for
this configuration there exist several regular vacuum states with a specific interpretation,
including a regular generalization of the Boulware state for a relativistic
field in a black hole geometry. In this way, we show that the freedom in
choosing a vacuum state is larger in dispersive theories than in relativistic
theories. This vacuum choice has crucial importance in issues such as the
presence or not of Hawking radiation.

This paper is organized as follows. Section II is devoted to the Bogolubov
approach and the definition of the appropriate inner product. A mode
analysis is also performed. Section III introduces the hydrodynamic
representation that leads to a Klein-Gordon equation in the long
wavelength regime and to its generalization for all wavelengths. The corresponding
generalized Klein-Gordon product is also introduced. In Section IV, we
show that both formalisms are actually equivalent and lead to the same
concept of positive and negative norm solutions. After defining the
appropriate inner product, Section V is devoted to a discussion of the different
vacuum state choices. We focus on a configuration of particular relevance for analogue gravity experiments: a one-dimensional stationary acoustic black hole. We conclude with some final comments in Section VI.

\section{Bogolubov approach}

Let us consider a condensed dilute gas of interacting bosons described in
terms of quantum field operators $\hat\psi$ and $\hat\psi^\dag$ that
annihilate and create particles (see e.g.\ Refs.~\cite{Fetter:1972,castin}).
The operator $\hat
\psi$ can be separated into two parts: a macroscopic wave function
$\psi_0$ describing the actual Bose-Einstein condensate and a quantum field operator $\hat \phi$ describing
perturbations around the condensate.

The order parameter $\psi_0$ satisfies the Gross-Pitaevskii equation
\begin{eqnarray}
i\hbar\partial_t\psi_0=\left(-\frac{\hbar^2}{2m}\nabla^2
+V_\text{ext}+ g |\psi_0^2|\right)\psi_0~,
\label{eq:gpeq}
\end{eqnarray}
where  $V_\text{ext}$ is an external potential, $m$ the atomic mass and $g$ the atomic
interaction constant (proportional to the $s$-wave scattering length).
We will also use the Madelung representation
\begin{eqnarray}
\psi_0=\sqrt{n_0}
e^{i\theta_0/\hbar}~,
\end{eqnarray}
 in terms of the number density of atoms $n_0$
and the phase $\theta_0$ of the condensate, which defines the flow
velocity potential (such that the flow velocity is ${\mathbf
v}=\nabla\theta_0/m$). Then Eq.\ (\ref{eq:gpeq}) translates into
\begin{eqnarray}
\hspace{-1.5ex}\partial_t n_0 +\nabla \cdot(n_0\nabla\theta_0)/m&=&0~,
\label{eq:continuity}\\
\hspace{-1.5ex}\partial_t\theta_0+\frac1{2m}(\nabla\theta_0)^2+V_\text{ext}+   g  n_0
-\frac{\hbar^2}{2m}\frac{\nabla^2\sqrt{n_0}}{\sqrt{n_0}}&=&0~,
\label{eq:bernoulli}
\end{eqnarray}
which are the continuity equation and the Bernoulli equation plus a
quantum potential term, respectively.

\subsection{Bogolubov equation}

The quantum perturbation field $\hat \phi$ satisfies the Bogolubov
equation (see e.g.\ Refs.\ \cite{Fetter:1972,castin})
\begin{eqnarray}
 i\hbar \partial_t \hat\phi
  =\mathcal H\hat\phi
  + mc^2e^{2i\theta_0/\hbar}\hat\phi^\dag~,
\label{eq:lingp}
\end{eqnarray}
where $c^2= g  n_0/m$ is the square of the local comoving speed of
sound and $\mathcal H$ is the operator
\begin{eqnarray}
\mathcal H=-\frac{\hbar^2}{2m}\nabla^2+
\frac{\hbar^2}{2m}\frac{\nabla^2c}{c}
 -\frac1{2}m{\mathbf v}^2-\partial_t\theta_0+mc^2~.
\end{eqnarray}
The commutation relation for the atomic creation and annihilation
operators
\begin{eqnarray}
[\hat\psi(\mathbf x,t),\hat\psi^\dag(\mathbf x',t)]=
\delta(\mathbf x-\mathbf x')
\end{eqnarray}
 translates into the commutation relation
\begin{eqnarray}
[\hat\phi(\mathbf x,t),\hat\phi^\dag(\mathbf x',t)]=\delta(\mathbf
x-\mathbf x')
\label{eq:commutator}
\end{eqnarray}
 for the
perturbation field operators, which create or annihilate   atoms in the
noncondensed part and correspondingly annihilate or create them in the
condensed phase. In other words, $\hat
\phi$ moves an atom from the noncondensed part to the condensate, and vice versa for $\hat\phi^\dag$.  This commutation relation is valid for condensed systems in
which the number of noncondensed atoms is very small compared to
the number of condensed ones.

Note that the Bogolubov equation~(\ref{eq:lingp}) could also have been obtained by expanding to
first order in $\phi$ the Gross-Pitaevskii equation~(\ref{eq:gpeq}) for
the order parameter $
\sqrt{n_0}e^{i\theta_0/\hbar}+\phi$.
In other words, the classical perturbation $\phi$ of the mean-field wave
function of the condensate satisfies exactly the same evolution equation as the
quantum fluctuations $\hat \phi$ around the condensed phase. From now
on, we will therefore drop the hat from the operator $\hat \phi$, unless necessary.

It is important to note that, although this equation can be obtained by
`linearizing' the Gross-Pitaevskii equation, it is a complex equation for a
genuinely complex field and is therefore non-linear: If $ \phi$ is a
solution, then in general $\alpha \phi$ is not (unless $\alpha$ is real).
Therefore, we cannot directly perform a mode expansion to find the
general solution. There exists a procedure~\cite{Fetter:1972,castin} which allows to overcome this problem by
enlarging the space in which we look for solutions to Eq.\ (\ref{eq:lingp})
and to define an inner product in this enlarged space.

 With this aim let us introduce the spinor field
\begin{eqnarray}
 \Phi=\frac1{\sqrt 2}\begin{pmatrix} \phi  \\ \tilde \phi  \end{pmatrix}~,
\end{eqnarray}
 subject to the evolution equation
\begin{eqnarray}
i\hbar\partial_t \Phi=\mathcal M \Phi~,
\label{eq:evolm}
\end{eqnarray}
where $\mathcal M$ is the operator
\begin{eqnarray}
\mathcal M&
           =&
           \left(
             \begin{array}{cc}
               \mathcal H&  mc^2e^{2i\theta_0/\hbar} \\
               - mc^2e^{-2i\theta_0/\hbar} & - \mathcal H \\
             \end{array}
           \right)\nonumber\\&
           =&\mathcal H\sigma_z
           +mc^2e^{2i\theta_0/\hbar} \sigma_+-mc^2e^{-2i\theta_0/\hbar} \sigma_-~.
\label{eq:enlargedbogol}
\end{eqnarray}
In this equation, $\sigma_\pm=(\sigma_x\pm
i\sigma_y)/2$ and $\sigma_{x,y,z}$ are the Pauli matrices.

This equation is now linear, i.e., if $\Phi$ is a solution, then so is $\alpha \Phi$
for any complex constant $\alpha$. The solutions to the Bogolubov
equation (\ref{eq:lingp}) are obtained by restricting the solutions of
Eq.\ (\ref{eq:evolm}) by the condition
\begin{eqnarray}
 \phi^*= {\tilde \phi}~,\quad \text{i.e.,}\quad \sigma_x\Phi^*=\Phi~.
\label{eq:physicalphi}
\end{eqnarray}

\subsection{Bogolubov inner product}

Taking into account that $\hat\phi$ has been defined as an operator
that annihilates atoms in the noncondensed part of the gas (and hence
creates them in the condensed part),  the  expectation value of
$\hat\phi^\dag\hat\phi$ will provide the number of noncondensed atoms
$N_1$ (under the assumption that this number is small compared to
the total number of atoms, as mentioned above). This condition translates
into the following normalization condition for the spinor $\Phi$:
\begin{eqnarray}
\langle \Phi,\Phi\rangle=N_1
\end{eqnarray}
in the standard inner product
\begin{eqnarray}
\langle\Phi|\Phi'\rangle=\int\text{d}^D\hspace{-0.4ex}x\,\Phi^\dag\Phi'
=\frac12\int\dDx(\phi^*\phi'+\tilde\phi^*\tilde\phi')~,
\label{eq:innermalo}
\end{eqnarray}
where $D$ is the number of spatial dimensions under
consideration. Since there is a continuous exchange of atoms between the
condensed and noncondensed phases, this norm will not be conserved in
time. Indeed, this fact is already encoded in the evolution equation
(\ref{eq:evolm}). Actually, it is straightforward to see that
\begin{eqnarray}
i\hbar\frac{\text{d}}{\text{d} t}\langle\Phi,\Phi'\rangle=
\langle (\mathcal M^\dag-\mathcal M)\Phi,\Phi'\rangle~.
\end{eqnarray}
However, the operator $\mathcal M$ is not selfadjoint in the
positive definite inner product (\ref{eq:innermalo}), but satisfies the
following properties:
\begin{eqnarray}
\sigma_x\mathcal M\sigma_x=-\mathcal M^*~,
\qquad
\sigma_z\mathcal M\sigma_z=\mathcal M^\dag~,
\label{eq:propm}
\end{eqnarray}
and hence this inner product is not preserved in the evolution, as we had
advanced.

These are the relevant properties of $\mathcal M$. In fact, in view of
the last equality, we can introduce a ``Bogolubov'' inner product
\begin{eqnarray}
\langle\Phi|\Phi'\rangle_\textsc{b}=\int\dDx\Phi^\dag\sigma_z\Phi'~,
\end{eqnarray}
in which $\mathcal M$ is selfadjoint. Indeed, it is straightforward to
check that
\begin{eqnarray}
\langle\Phi|\mathcal M\Phi'\rangle_\textsc{b}=\langle\mathcal M\Phi|\Phi'\rangle_\textsc{b}~.
\end{eqnarray}
Therefore we see that the price to pay for making the
evolution operator $\mathcal M$ selfadjoint is the introduction of an
inner product $\langle\cdot|\cdot\rangle_\textsc{b}$ which is not
positive definite. Indeed, this Bogolubov inner product has the following properties:
\begin{itemize}
\item It is conserved in the evolution of the lab time $t$.
\item It is hermitian, i.e., $\langle\Phi|\Phi'\rangle_\textsc{b}^*=\langle\Phi'|\Phi\rangle_\textsc{b}$.
\item It is antilinear
in the first argument and linear in the second, i.e., for any complex
number $\alpha$,
\begin{eqnarray}
\langle\alpha\Phi|\Phi'\rangle_\textsc{b}&=&\alpha^*\langle\Phi|\Phi'\rangle_\textsc{b}~,\quad
\langle\Phi|\alpha\Phi'\rangle_\textsc{b}=\alpha\langle\Phi|\Phi'\rangle_\textsc{b}~.
\end{eqnarray}
\item  It is not
positive definite, since it satisfies
\begin{eqnarray}
\langle\sigma_x\Phi^*|\sigma_x\Phi'^*\rangle_\textsc{b}=
-\langle\Phi'| \Phi \rangle_\textsc{b}~.
\label{eq:bogolnonpos}
\end{eqnarray}
\end{itemize}

Finally, note that the physical solutions, i.e.\ those that satisfy $\sigma_x\Phi^*=\Phi$
because of condition (\ref{eq:physicalphi}),
have zero norm, as can easily be seen from Eq.\ (\ref{eq:bogolnonpos}).

\subsection{Mode expansion}

The evolution operator $\mathcal M$ is selfadjoint in a non-positive-definite inner product and therefore it may have complex eigenvalues. We
will assume that the condensate is stable, which implies that genuinely
complex frequencies cannot be present.

In view of the properties (\ref{eq:propm}), it is easy to see that, if
\begin{eqnarray}
U_k=\frac1{\sqrt 2}\begin{pmatrix} u_k  \\ v_k   \end{pmatrix}
\end{eqnarray}
 is an eigenspinor of
$\mathcal M$ with eigenvalue $\omega_k$, i.e., if
\begin{eqnarray}
\mathcal M
U_k=\omega_kU_k~,
\end{eqnarray}
 then
 \begin{eqnarray}
 V^*_k=\sigma_x U_k^*=\frac1{\sqrt 2}
\begin{pmatrix} v_k^*  \\   u_k^*  \end{pmatrix}
 \end{eqnarray}
is an eigenspinor of $\mathcal M$ with eigenvalue $-\omega_k$. Besides,
$\sigma_z U_k $ is an eigenvector of $\mathcal M^\dag$ with
eigenvalue $\omega_k$. Furthermore, the modes $U_k$ and $V^*_k$
are orthogonal (and can be chosen orthonormal) in the Bogolubov inner
product:
\begin{eqnarray}
\langle U_k|U_l\rangle_\textsc{b}
&=&\frac12\int\dDx (u_k^*u_l-v_k^*v_l)=\delta_{kl}~,\\
\langle U_k|V^*_l\rangle_\textsc{b}
&=&\frac12\int\dDx (u_k^*v_l^*-v_k^*u_l^*)=0~,\\
\langle V^*_k|V^*_l\rangle_\textsc{b}
&=&\frac12\int\dDx (v_kv_l^*-u_ku_l^*)=-\delta_{kl}~.
\end{eqnarray}
Any spinor $\Phi$, solution to Eq.\ (\ref{eq:evolm}), can be
expanded in this basis:
\begin{eqnarray}
\Phi=\sum_k(a_k U_k+ b_k^* V_k^*)~,
\end{eqnarray}
and its norm is given by
\begin{eqnarray}
\langle\Phi|\Phi\rangle_\textsc{b}=\sum_k(|a_k|^2-|b_k|^2)~.
\end{eqnarray}
We therefore again see that the physical solutions---those satisfying the
condition (\ref{eq:physicalphi})---have zero norm, since they satisfy
$a_k=b_k$. Note that the modes themselves are not physical, not only
because they may be generalized eigenvectors normalized to the Dirac delta,
but also because in general they do not satisfy Eq.\ (\ref{eq:physicalphi}):
$\sigma_xU_k^*=V_k^*\neq U_k$, as we have seen.

Finally, the number of field degrees of freedom carried by the spinor
$\Phi$ is just two. Indeed, two complex (four real) functions are needed
at an initial time to obtain the value of $\Phi$ at any other time.
Condition (\ref{eq:physicalphi}), which ensures the physical nature of the
configuration, reduces this number to one (two real initial functions),
which is precisely the  number of field degrees of freedom of a real
relativistic scalar field. In the next section we will actually describe the
condensate perturbations as a real scalar field satisfying (in
the appropriate limit of long wavelengths) a relativistic wave equation.

\section{Klein-Gordon approach}

As an alternative to the approach followed in the previous section, we
can linearize the continuity (\ref{eq:continuity}) and Bernoulli
(\ref{eq:bernoulli}) equations around a background condensate
characterized by $n_0$ and $\theta_0$. Let us introduce the density
$\tilde n_1$ and phase $\theta_1$ perturbations:
\begin{eqnarray} n=n_0+ g ^{-1}\tilde{n}_1~,\qquad
\theta=\theta_0+\theta_1~.
\end{eqnarray}
These perturbations obey the equations
\begin{eqnarray}
&&
\partial_t \tilde{n}_1 + \nabla \cdot( \tilde{n}_1 {\mathbf v} + c^2 \nabla \theta_1)=0~,
\label{eq:hyd1}\\
&&
\partial_t \theta_1  + {\mathbf v} \cdot \nabla \theta_1
+(1-\Theta)\tilde{n}_1=0~,
\label{eq:hyd2}
\end{eqnarray}
where  $\Theta$ is the operator
\begin{eqnarray}
\Theta=\frac14\xi^2
\nabla[c^2\cdot\nabla (c^{-2}\star)]~,
\end{eqnarray}
$\xi=
\hbar/(mc)$ is the healing length, and
the $\star$ stands for the argument upon which $\Theta$ acts.

Note that the perturbation fields $\phi$, $\tilde\phi$ introduced in the
previous sections are related to the complexified density $\tilde n_1$
and phase $\theta_1$    perturbations in the following way:
\begin{eqnarray}
\phi&=&e^{i\theta_0/\hbar}\frac{1}{\sqrt{gm}}\left( \frac{1}{ 2 c} {\tilde n}_1 +
i \frac{1}{\xi} \theta_1 \right)~,\nonumber\\
\tilde\phi&=&e^{-i\theta_0/\hbar}\frac{1}{\sqrt{gm}}\left( \frac{1}{ 2 c} {\tilde n}_1
- i \frac{1}{\xi} \theta_1 \right)~,
\label{eq:phintheta}
\end{eqnarray}
in terms of which
\begin{eqnarray}
&&{\tilde n}_1 = c \sqrt{gm} (e^{-i\theta_0/\hbar} \phi
 + e^{i\theta_0/\hbar} \tilde\phi)~,
\nonumber\\
&&\theta_1=
 -i\frac {\xi}{2} \sqrt{gm} (e^{-i\theta_0/\hbar} \phi
  - e^{i\theta_0/\hbar} \tilde\phi)~.
  \label{eq:phinthetainv}
\end{eqnarray}
The condition (\ref{eq:physicalphi}) that $\phi$ and $\tilde\phi$
represent a physical solution to the Bogolubov equation (\ref{eq:lingp})
translates into reality conditions for $\tilde n_1$ and $\theta_1$. It is
also interesting to note that the commutation relation for these two
fields is
\begin{eqnarray}
[\hat{\tilde n}_1(\mathbf x,t),\hat\theta_1(\mathbf x',t)] =ig\hbar
\delta(\mathbf x-\mathbf x') ~,
\end{eqnarray}
as a direct consequence of the commutation relation
(\ref{eq:commutator}) for~$\hat\phi$. Thus, $\tilde n_1$ and $\theta_1$
are canonically conjugate fields.

\subsection{Generalized Klein-Gordon equation}

We can now combine  the two  equations (\ref{eq:hyd1}) and
(\ref{eq:hyd2}) for $\theta_1$ and $\tilde{n}_1$ in order to obtain a
second order (in time) differential equation for $\theta_1$. More
explicitly, we can obtain $\tilde{n}_1$ as a function of $\theta_1$ from
Eq.\ (\ref{eq:hyd2}) by formally inverting the operator $(1-\Theta)$:
\begin{eqnarray}
\tilde{n}_1=-\mathcal W(\partial_t+{\mathbf v}\cdot \nabla)\theta_1~,
\label{eq:ntheta}
\end{eqnarray}
where
\begin{eqnarray}
\mathcal W
=(1-\Theta)^{-1}=\sum_{n=0}^\infty \Theta^n~.
\end{eqnarray}
Note that generally this is a well-defined procedure because of  the
negativity of the  operator $\Theta$. Indeed, for homogeneous profiles,
$\Theta$ is  proportional to the Laplacian, which is obviously a negative
operator. As long as the profile is sufficiently smooth such that $c$ is slowly varying on the healing length scale, this negative character will not be altered. Relevant departures from this behaviour would require profiles whose density varies significantly within length scales comparable with the healing length, which is not only easily avoidable in practice but might actually be very hard to realize. Keeping these comments in mind,  we can insert Eq.\ (\ref{eq:ntheta}) into Eq.\ (\ref{eq:hyd1}) to obtain the single equation
\begin{eqnarray}
\hspace{-1ex}-[\partial_t+ \nabla\cdot ({\mathbf v}\star)]\mathcal W
(\partial_t+{\mathbf v}\cdot\nabla)\theta_1+\nabla\cdot(c^2\nabla\theta_1)=0~.
\end{eqnarray}
This equation is a higher order differential equation which generalizes the
Klein-Gordon equation.

In the limit $\mathcal W \to 1$, attained when the gradients in
$\theta_1$ are relevant only for length scales much larger than the healing length,  a
proper Klein-Gordon equation is recovered. Indeed, in this case, this
evolution equation can be written in the form~\cite{Barcelo:2005fc}
\begin{eqnarray}
\partial_\mu(\sqrt{-g}g^{\mu\nu}\partial_\nu\theta_1)=0~,
\end{eqnarray}
where $g_{\mu\nu}$ is the acoustic  metric
\begin{eqnarray}
g_{\mu\nu}=c^{{2}/{(D-1)}}\left(
                      \begin{array}{cc}
                       -(c^2-{\mathbf v}^2) & -{\mathbf v}^{\textsc{t}} \\
                        -{\mathbf v} & \openone \\
                      \end{array}
                    \right)~.
\end{eqnarray}
The corresponding Klein-Gordon inner product can be written as
\begin{eqnarray}
\langle\theta_1|\theta_1'\rangle_{\textsc{kg}}&=&
i\int\dDx \sqrt
q\theta_1^*\overleftrightarrow{\partial_n}\theta_1'\nonumber\\
&=&i\int\dDx
\theta_1^*\overleftrightarrow{ (\partial_t+{\mathbf v}\cdot\nabla) }
\theta_1'~,
\label{eq:kgprod}
\end{eqnarray}
where $\sqrt{q}=c^{D/(D-1)}$ is the determinant of the metric in the
spatial slice $t$=constant, $n^\mu=c^{-D/(D-1)}(1,{\mathbf v})$ is its
normal, and $\partial_n=n^\mu\partial_\mu$.

\subsection{Generalized Klein-Gordon product}

In the general case, we can introduce an inner product that generalizes
the Klein-Gordon product by taking into account that the time derivative
term is now modified by the operator $\mathcal W$. In $D$ spatial
dimensions, this $\mathcal W$-Klein-Gordon inner product turns out to
be
\begin{eqnarray}
\langle\theta_1|\theta_1'\rangle_{\mathcal W\text{\!-}\textsc{kg}}=
i\int\dDx
\theta_1^*\overleftrightarrow{[\mathcal W(\partial_t+{\mathbf v}\cdot\nabla)]}
\theta_1'~.
\label{eq:w-kgprod}
\end{eqnarray}
 It is clear from this expression that it reduces to the standard relativistic
Klein-Gordon product in the limit $\mathcal W\to 1$.

It should be stressed that the operator $\mathcal W$ breaks the local
Lorentz invariance of the Klein-Gordon equation. This means that an effective curved spacetime geometry is recovered only for condensate perturbations such that $\mathcal W$ is very close to $1$. Furthermore, as we will see below, the dispersion relation for this $\mathcal W$-Klein-Gordon
equation is the same as that for the Klein-Gordon equation modified with
fourth-order spatial derivatives. Here, however, the Klein-Gordon inner product
is modified in each $t=\text{constant}$ slice by the action of the
operator $\mathcal W$. The Klein-Gordon equation modified with
fourth-order spatial derivatives, on the other hand, shares the Klein-Gordon product in each $t=$constant slice (\ref{eq:kgprod}) with the proper Klein-Gordon equation~\cite{Corley:1998rk}.

In spite of these modifications, the $\mathcal W$-Klein-Gordon product
(\ref{eq:w-kgprod}) shares the following properties with the standard
Klein-Gordon product:
\begin{itemize}
\item It is conserved  in the lab time $t$.
However, unlike the Klein-Gordon product, which is conserved in any
inertial time, the $\mathcal W$-Klein-Gordon product is conserved only
in the lab time. This is a logical consequence of the fact that local Lorentz invariance is
no longer in operation.

\item It is hermitian, i.e.,
$\langle\theta_1|\theta_1'\rangle_{\mathcal
W\text{\!-}\textsc{kg}}^*
=\langle\theta_1'|\theta_1\rangle_{\mathcal
W\text{\!-}\textsc{kg}}$.

\item It is antilinear in the first argument and linear in the second.
\item It is not positive definite. Indeed,
\begin{eqnarray}
\langle\theta_1^*|\theta_1'^*\rangle_{\mathcal
W\text{\!-}\textsc{kg}}
=-\langle\theta_1'|\theta_1\rangle_{\mathcal
W\text{\!-}\textsc{kg}}~.
\end{eqnarray}
\item There exists at least one basis $\{\theta_{1i}\}$ of orthonormal solutions such that
\begin{eqnarray}
\langle\theta_{1i}|\theta_{1j}\rangle_{\mathcal W\text{\!-}\textsc{kg}}&=&\delta_{ij}~,
\nonumber\\
\langle\theta_{1i}|\theta_{1j}^*\rangle_{\mathcal W\text{\!-}\textsc{kg}}&=&0~,\label{eq:ortho-rel}
\\
\langle\theta_{1i}^*|\theta_{1j}^*\rangle_{\mathcal W\text{\!-}\textsc{kg}}&=&-\delta_{ij}~.
\nonumber\end{eqnarray}
\item The norm of any solution $\theta_1=\sum_i(a_i\theta_{1i}+b_i^*\theta_{1i}^*)$ is
\begin{eqnarray}
\langle\theta_1|\theta_1\rangle_{\mathcal W\text{\!-}\textsc{kg}}=\sum_i(|a_i|^2-|b_i|^2)~.
\end{eqnarray}
This norm vanishes for real scalar fields such as the
physical phase perturbation.
\end{itemize}

At this stage, it is worth noting that the $\mathcal W$-Klein-Gordon
inner product has the same properties that we stated above for the
Bogolubov inner product. In fact we will see that these two products are
indeed equivalent.

\subsection{Mode expansion}

We will now discuss the form of the modes and their normalization. For the sake of simplicity and definiteness, let us concentrate on a case
of particular interest, namely, the case of a background profile which
becomes time-independent in the asymptotic future. Then we can try an
ansatz of the form $Ae^{-i\omega u(t,\mathbf x)}$, which in the pure
one-dimensional Klein-Gordon case leads to  exact orthonormal modes
$(2c|k|)^{-1/2}e^{-i(\omega t-kx)}$.

 Let us define
\begin{eqnarray}
\bar\omega(t,\mathbf x)=\omega\partial_tu~,\quad
\mathbf k(t,\mathbf x)=-\omega\nabla u~.
\end{eqnarray}
The condition that the profile becomes stationary in the asymptotic
future implies that $\bar\omega(t\to\infty,\mathbf x)=\omega$ and
that $\mathbf k(t,\mathbf x)$ is time-independent in this limit. Then $A
e^{-i\omega u}$ is an approximate solution which can be found in the
regime where $\mathbf k$, $\bar\omega$, and $A$ are slowly varying
functions (in space and time). Introduction of this approximate solution into the modified Klein-Gordon equation yields, to  lowest order, the dispersion
relation
\begin{eqnarray}
(\bar\omega-{\mathbf v}\cdot\mathbf k)^2=c^2\mathbf k^2 \Gamma_{\mathbf
k}^2~,
\label{eq:disp-rel}
\end{eqnarray}
and hence the form of $u(t,\mathbf x)$, where
\begin{eqnarray}
\Gamma_{\mathbf k}=+\sqrt{1+\xi^2 \mathbf k^2/4}~.
\end{eqnarray}
The next-to-lowest order yields the prefactor $A$ so that a complete
set of approximate modes is given by
\begin{eqnarray}
\theta_{1\mathbf k}=\frac{\sqrt{\Gamma_{\mathbf
k}}}{\sqrt{4\pi c|\mathbf k|}}e^{-i\omega  u}~.
\end{eqnarray}
The next order provides information about the spread of these modes in
the directions perpendicular to that of group propagation. The relativistic
limit $\mathcal W\to 1$ obviously corresponds to $\Gamma_{\mathbf
k}\to 1$, i.e., to the long wavelength limit $\xi |\mathbf k|\ll1$, which
gives rise to the well-known Klein-Gordon modes. These approximate modes
are orthonormal (in the same level of approximation) in the $\mathcal
W$-Klein-Gordon product (\ref{eq:w-kgprod}) as can be seen by evaluating
it at $t\to\infty$. Evaluated at any other finite time $t$, small
deviations which are consistent with the level of approximation that we
are using may appear.

As mentioned above, any real  solution $\theta_1$ can be written as a
linear combination of these modes
\begin{eqnarray}
\theta_1=\sum_{\mathbf k} (a_{\mathbf k}\theta_{1\mathbf k}
+a_{\mathbf k}^*\theta_{1\mathbf k}^*)
\end{eqnarray}
and therefore has zero norm.

For completeness, it is also straightforward to check from Eq.\ (\ref{eq:ntheta}) that
\begin{eqnarray}
\tilde n_{1\mathbf k}=
i\frac{c|\mathbf k|}{\Gamma_{\mathbf k}}\theta_{1\mathbf k}~.
\label{eq:nthetak}
\end{eqnarray}

Finally, before examining the relation between the Bogolubov and the Klein-Gordon formalisms, we mention the existence of a third alternative approach \cite{Macher:2009nz}, which leads to a genuinely complex differential equation of second order in time and fourth order in space for the field perturbation $\phi$. Indeed, if we write $\phi^*$ in terms of $\phi$ from the Bogolubov equation and use this expression in the complex conjugate of this same equation, we obtain
\begin{align}
\Big\{\left[ \hbar \left(\partial_t + \mathbf v \cdot\nabla \right) -i
T_\rho\right]\frac{1}{c^{2}} \left[ \hbar \left(\partial_t + \mathbf v
\cdot\nabla \right)+ i T_\rho\right]& \nonumber\\
{}+ 2m T_\rho \Big\} \varphi = 0&~,
\end{align}
where $\varphi=\phi/\psi_0$ and $T_\rho=-\frac{\hbar^2}{2mc^2}\nabla
\cdot(c^2\nabla\star)$.
It might seem at first sight that the number of degrees of freedom is equivalent to that of a complex scalar field (i.e., two sets of complex Fourier coefficients). However $\varphi$ and $\varphi^*$ are not independent but are related by the Bogolubov equation itself, leaving---as before---just one field degree of freedom (a single set of complex Fourier coefficients) and, in fact, the scalar product in the space of solutions of this equation is the Bogolubov one.

\section{Equivalence between both products}

The following question of interest regards the relation between the Bogolubov product
 and the $\mathcal W$-Klein-Gordon product introduced in the previous sections. More specifically,
what is the relation between the concepts of quasiparticle and antiquasiparticle
in both formalisms?

Let us consider two spinors $\Phi$ and $\Phi'$, solutions to the
Bogolubov evolution equation (\ref{eq:evolm}). Then, taking into account
the relation (\ref{eq:phintheta}) between both representations, i.e.,
between the perturbation fields $\phi$, $\tilde\phi$  and $\tilde{n}_1$,
$\theta_1$, we obtain
\begin{eqnarray}
\langle \Phi|\Phi'\rangle_\textsc{b}
&=&\frac12\int\dDx (\phi^*\phi'-\tilde\phi^*\tilde\phi')
\nonumber\\
&=&\frac{i}{ 2g \hbar}\int \dDx(\tilde{n}_1^*\theta_1'-\theta_1^*\tilde{n}_1')~.
\end{eqnarray}

Finally, the relation (\ref{eq:ntheta}) between the density and phase
perturbation allows us to write this Bogolubov product as
\begin{eqnarray}
\langle \Phi|\Phi'\rangle_\textsc{b}
&=&\frac{i}{2g\hbar}\int\dDx
\theta_1^*\overleftrightarrow{[\mathcal W(\partial_t+{\mathbf v}\cdot\nabla)]}
\theta_1'\\
&=&\frac{1}{ 2g \hbar}\langle\theta_1|\theta_1'\rangle_{\mathcal
W\text{\!-}\textsc{kg}}~.
\end{eqnarray}
So, we see that the Bogolubov and the $\mathcal W$-Klein-Gordon
products are indeed equivalent.

Also, provided a set of orthonormal modes $\theta_{1\mathbf k}$ of the
\hbox{$\mathcal W$-Klein}-Gordon equation, it is straightforward to construct
an orthonormal set of modes for the Bogolubov equation  (\ref{eq:evolm})
by means of the relations (\ref{eq:phintheta}) and (\ref{eq:nthetak}):
\begin{eqnarray}
u_{\mathbf k}&=&e^{i\theta_0/\hbar}\frac{1}{\sqrt{gm}\xi}
\left( 1+\frac{\xi|\mathbf k|}{2\Gamma_{\mathbf k}} \right)
\theta_{1\mathbf k}~,\nonumber\\
v_{\mathbf k}&=&e^{-i\theta_0/\hbar}\frac{1}{\sqrt{gm}\xi}
\left( -1+\frac{\xi|\mathbf k|}{2\Gamma_{\mathbf k}} \right)
\theta_{1\mathbf k}~.
\end{eqnarray}
Therefore positive (resp.\ negative) norm modes in the \hbox{$\mathcal
W$-Klein}-Gordon product are mapped to positive (resp.\ negative) norm
modes in the Bogolubov product and vice versa. This means that, whether
we choose to analyze quasiparticle creation processes (e.g., Hawking
radiation in a black hole configuration) in a modified relativistic
framework such as in~\cite{Barcelo:2008qe} or a condensed-matter context such as in~\cite{Macher:2009nz}, the results should coincide. The
reason for this coincidence is that, as we have seen, there is a
one-to-one relation between both formalisms, and the concepts of
quasiparticle and antiquasiparticle, as well as the ambiguities inherent 
in these definitions (which are related to the observer-dependence of the
concepts involved), are the same in both formalisms.

\section{On the choice of vacuum state}

One can now proceed with the quantization of the system following the
standard Fock procedure. One only needs to find an orthonormal mode
basis to expand the field operators that characterize the quantum
perturbations. In this manner, one can define creation and annihilation
operators and a vacuum state for the system. The fact that the inner
product is not positive definite tells us that the selection of a specific set of positive (negative) norm modes can be done in many different ways. To illustrate the procedure, let us consider a
one-dimensional stationary flow in the condensate which simulates the
presence of a black hole, but with the internal singularity substituted by a
second asymptotic region~\cite{Barcelo:2004wz}. Among the different
sets of modes that can be selected, there are three of special relevance, which
we will call the ``in'' set, the ``out'' set and the ``stationary'' set,
whose meaning will be discussed shortly. Each positive energy mode of
each set can be characterized by its frequency (which is invariant due to
the stationarity of the system) and by a discrete label which, depending
on the frequency, can acquire the values 1, 2 or 1, 2, 3 (see the discussion in~\cite{Macher:2009tw} regarding the number of normalizable modes in different 
configurations). 

We will now classify these modes, using the following subscripts. The $p$
and $f$ subscripts represent ``past'' and ``future'' (in a scattering
process) while $u$ and $w$ represent the ``right-going'' and
``left-going'' character of wave packets  centered   around the particular mode
frequency with respect to the lab.

\begin{figure}
\includegraphics[width=.40\textwidth]{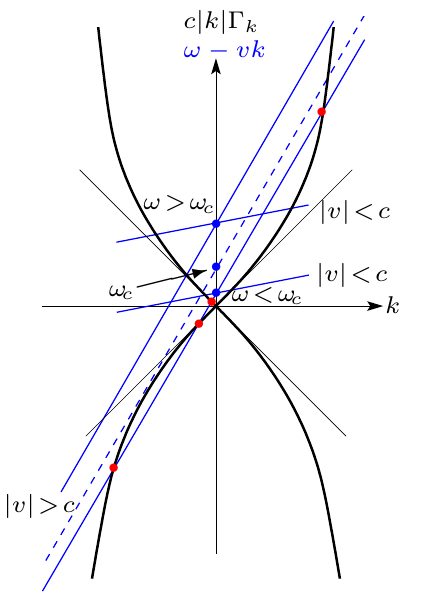}
\bigskip
\caption{\label{Fig:disp-rel} 
Dispersion relation~(\ref{eq:disp-rel}) for various values of $\omega$ and of $|v|$, scaled at $c=1$. The intersection points between the line $\omega-vk$ (in blue) and the various branches of the dispersion relation (in black) mark the real (normalisable) mode solutions for a given $\omega$. The critical frequency $\omega_c$ represents the frequency at which additional, `extraordinary' roots appear for the maximal $|v|$ attained in the configuration. Then, for $\omega>\omega_c$, there are always only two normalisable solutions. For $\omega<\omega_c$, there can be either two or four normalisable solutions, depending on the value of $|v|$.
}
\end{figure}

Let us consider the dispersion relation~(\ref{eq:disp-rel}) formally as a relation between $\omega$, 
$k$ and $x$ (through $v(x)$). Then, for any given  real frequency  $\omega$ and any point $x$ in the configuration, the
dispersion relation may have two or four real roots $k$, see Fig.~\ref{Fig:disp-rel}. Actually, there exists  a critical frequency $\omega_c$ such that, for
$\omega>\omega_c$, independently of $x$, there  always exist two real
roots only, one positive and one negative. We denote by $k_{up}$ the
positive root when $x\to-\infty$, by $k_{wp}$ the absolute value of the
negative root when $x\to+\infty$, by $k_{uf}$ the positive root when
$x\to+\infty$ and finally by $k_{wf}$ the absolute value of the negative
root when $x\to-\infty$. For $0<\omega<\omega_c$, depending on $x$,
there can be either two or four roots (there is a critical position at which there
are just three solutions; here we will not discuss this critical situation).
Two of the roots are always equivalent to the previous ones; we will use
for them the same notation. In case of the existence of four real roots, the two additional ones always correspond to negative values of $k$. We will call these additional modes `extraordinary' because they are absent in the subsonic regime. 
Let us then denote by $k_{ep}$ the absolute value of the most negative  extraordinary root when
$x\to-\infty$ and by $k_{ef}$ the absolute value of the second most
negative extraordinary root when $x\to-\infty$. The corresponding wave packets are right-going ($k_{ep}$) and left-going 
($k_{ef}$), respectively.

We summarize this classification in table~\ref{tb:modes}.

\begin{table}
\begin{tabular}{|l|c|c|c|c|}
\hline
& \multicolumn{2}{|c|}{~$t \to -\infty$~}  & \multicolumn{2}{|c|}{~$t \to +\infty$~} \\
\cline{2-5}
& $\Longrightarrow$ & $\Longleftarrow$ & $\Longrightarrow$ & $\Longleftarrow$ \\
\hline
~$x\to +\infty$ ($v<c$) & & $k_{wp}$ & $k_{uf}$ & \\
\hline
~$x\to -\infty$ ($v>c$) and $\omega>\omega_c$~ & $k_{up}$ & & & $k_{wf}$\\
\hline
\multirow{2}{*}{~$x\to -\infty$ ($v>c$) and $\omega<\omega_c$~} & $k_{up}$ &  &  & $k_{wf}$\\
& $k_{ep}$ & & & $k_{ef}$\\
\hline
\end{tabular}
\caption{\label{tb:modes} Classification of normalizable mode solutions (real roots) in the asymptotic regions. The arrows indicate the left- or right-going character of the corresponding wave packets with respect to the lab.}
\end{table}

Given the equivalence displayed in the previous section, we are free to
describe the perturbations either by a spinor in the Bogolubov formulation
or by the phase perturbation operator $\hat\theta$ in the Klein-Gordon formulation. 
For notational simplicity, we omit the subscript $1$ for the perturbations in this section.

\subsection{``In'' vacuum state}
\begin{figure}
\includegraphics[width=.45\textwidth]{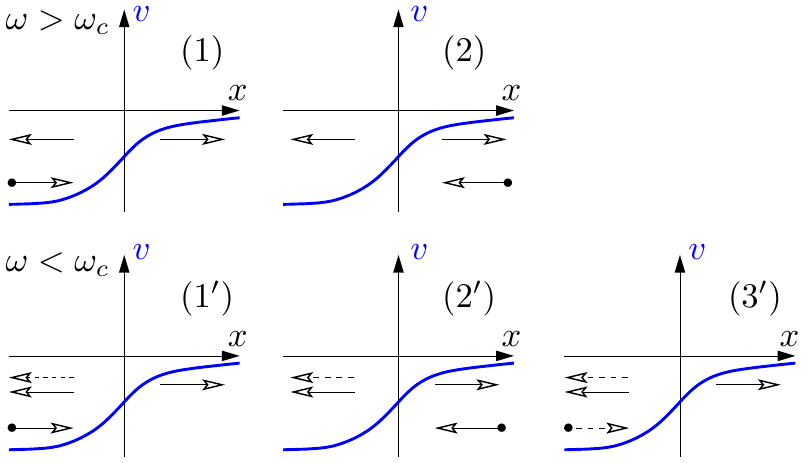}
\bigskip
\caption{\label{Fig:in-basis} ``In'' basis.
\underline{$\omega>\omega_c$:} $(1)$ represents a  scattering process in which
there is an incoming wave from the left; $(2)$ represents a scattering
process in which there is an incoming wave from the right.
\underline{$0<\omega<\omega_c$:} $(1')$ represents a  scattering process in which
there is an incoming wave from the left; $(2')$ represents a  scattering
process in which there is an incoming wave from the right; $(3')$ represents
a  scattering process in which there is an extraordinary incoming wave
from the left.
 Solid arrows represent ordinary modes while dashed arrows represent extraordinary modes.
}
\end{figure}

It can be shown~\cite{Macher:2009tw} that the positive energy modes
that constitute the ``in'' set (such that, in the ``in'' vacuum defined by
them, there are no quasiparticles in the asymptotic past) can be described as follows,
up to a global mode-dependent normalization constant:
\begin{itemize}
\item For $\omega>\omega_c$,
\begin{align}
\theta^{\rm in}_{\omega,1}
 &\stackrel{x\to-\infty}{\longrightarrow}
 e^{-i\omega t}\left(e^{  ik_{up} x} +
{\tilde R}^{\rm in}_\omega e^{  - i k_{wf} x}\right)~, \nonumber\\
\theta^{\rm in}_{\omega,1}
 &\stackrel{x\to+\infty}{\longrightarrow}
  e^{-i\omega t}\left({\tilde T}^{\rm in}_\omega
 e^{  i k_{uf} x}\right)~;
\\[.6\baselineskip]
\theta^{\rm in}_{\omega,2}
 &\stackrel{x\to+\infty}{\longrightarrow}
  e^{-i\omega t}\left(e^{  - ik_{wp} x} +
  R^{\rm in}_\omega e^{  i k_{uf} x}\right)~, \nonumber\\
\theta^{\rm in}_{\omega,2}
 &\stackrel{x\to-\infty}{\longrightarrow}
 e^{-i\omega t}\left(T^{\rm in}_\omega e^{  - i k_{wf} x}\right)~;
\end{align}

\item For $\omega<\omega_c$,
\begin{align}
\theta^{\rm in}_{\omega,1}
&\stackrel{x\to-\infty}{\longrightarrow}
  e^{-i\omega t}\left(e^{    ik_{up} x} +
{\tilde R}^{\rm in}_\omega e^{  - i k_{wf} x} +
 {\tilde R}^{\rm in}_{\omega,e} e^{  - i k_{ef} x}\right)~, \nonumber\\
\theta^{\rm in}_{\omega,1}&\stackrel{x\to+\infty}{\longrightarrow}
 e^{-i\omega t}\left({\tilde T}^{\rm in}_\omega
e^{    i k_{uf} x}\right)~; \\[.6\baselineskip]
\theta^{\rm in}_{\omega,2}
&\stackrel{x\to+\infty}{\longrightarrow}
 e^{-i\omega t}\left(e^{ - ik_{wp} x} +
R^{\rm in}_\omega e^{    i k_{uf} x}\right)~, \nonumber\\
\theta^{\rm in}_{\omega,2}
&\stackrel{x\to-\infty}{\longrightarrow}
 e^{-i\omega t}\left(T^{\rm in}_\omega e^{  - i k_{wf} x}+
T^{\rm in}_{\omega,e} e^{  - i k_{ef} x}\right)~; \\[.6\baselineskip]
\theta^{\rm in}_{\omega,3}
&\stackrel{x\to-\infty}{\longrightarrow}
 e^{i\omega t}\left(e^{  - ik_{ep} x} +
{\bar R}^{\rm in}_\omega e^{  i k_{wf} x} +
{\bar R}^{\rm in}_{\omega,e} e^{  i k_{ef} x}\right)~, \nonumber\\
\theta^{\rm in}_{\omega,3}
&\stackrel{x\to+\infty}{\longrightarrow}
 e^{i\omega t}\left({\bar T}_\omega e^{  - i k_{uf} x}\right)~;
\end{align}

\end{itemize}
where the  $R$'s and $T$'s are mode-dependent reflection and
transmission coefficients. In other words, every mode behaves in the specific form displayed by the above formulas in the appropriate asymptotic limits. These modes can be pictorially described  as
representing elementary scattering processes (see
Fig.~\ref{Fig:in-basis}). For $\omega>\omega_c$ there are two
independent and orthonormal scattering processes: $(1)$ represents the
scattering of an incoming wave from the left and $(2)$ the scattering of
an incoming wave from the right. For $0<\omega<\omega_c$ there are
three independent and orthonormal scattering processes: $(1')$ an
incoming wave from the left, $(2')$ an incoming wave from the right, and
$(3')$ an incoming extraordinary wave from the left. All these three
processes end up being a combination of three outgoing waves at future
infinity, one of them extraordinary.  Notice that $\omega$ in the
exponential of the mode  $\theta^{\rm in}_{\omega,3}$ appears with a
positive sign, contrarily to the negative sign in the other modes. 
These signs are chosen in order to define positive energy (norm) modes.\footnote{In the notation of Ref.~\cite{Macher:2009tw} our
$\theta^{\rm in}_{\omega,3}$ corresponds to 
$\varphi_{-\omega}^u$.}

\subsection{``Out'' vacuum state}
\begin{figure}
\includegraphics[width=.45\textwidth]{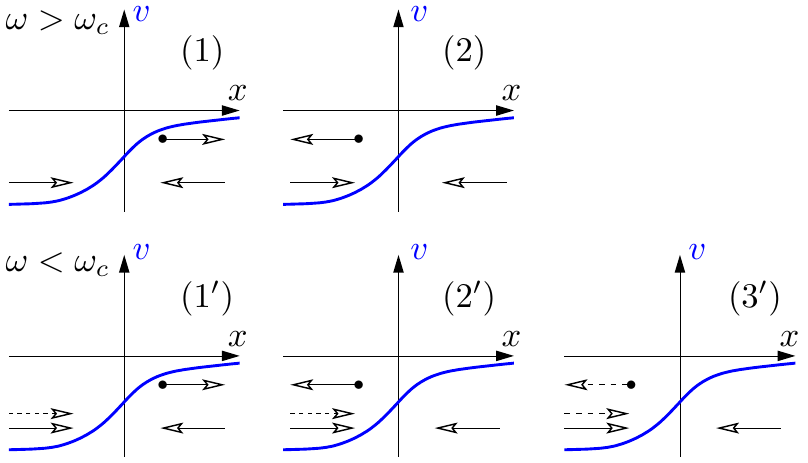}
\bigskip
\caption{\label{Fig:out-basis} ``Out'' basis. Interpretation as in (time-reversed) Fig. \ref{Fig:in-basis}.}
\end{figure}

In an equivalent way, one can construct the positive energy ``out''
basis~(such that, in the ``out'' vacuum defined by them, there are no
quasiparticles in the asymptotic future)~\cite{Macher:2009tw}, up to a global
mode-dependent normalization constant:
\begin{itemize}
\item For $\omega>\omega_c$,
\begin{align}
\theta^{\rm out}_{\omega,1}
&\stackrel{x\to+\infty}{\longrightarrow}  e^{-i\omega t}
\left(e^{  ik_{uf} x} + R^{\rm out}_\omega e^{  - i k_{wp} x}\right)~, \nonumber\\
\theta^{\rm out}_{\omega,1}
&\stackrel{x\to-\infty}{\longrightarrow}
  e^{-i\omega t}\left(T^{\rm out}_\omega e^{  i k_{up} x}\right)~;
 \\[.6\baselineskip]
\theta^{\rm out}_{\omega,2}
&\stackrel{x\to-\infty}{\longrightarrow}
  e^{-i\omega t}\left(e^{  - ik_{wf} x} +
 {\tilde R}^{\rm out}_\omega e^{  i k_{up} x}\right)~, \nonumber\\
\theta^{\rm out}_{\omega,2}
&\stackrel{x\to+\infty}{\longrightarrow}
  e^{-i\omega t}
 \left({\tilde T}^{\rm out}_\omega e^{  - i k_{wp} x}\right)~;
\end{align}
\item For $0<\omega<\omega_c$,
\begin{align}
\theta^{\rm out}_{\omega,1}
&\stackrel{x\to+\infty}{\longrightarrow}
  e^{-i\omega t}\left(e^{  ik_{uf} x} +
 R^{\rm out}_\omega e^{  - i k_{wp} x}\right)~, \nonumber\\
\theta^{\rm out}_{\omega,1}
&\stackrel{x\to-\infty}{\longrightarrow}
  e^{-i\omega t}\left(T^{\rm out}_\omega e^{  i k_{up} x}+
 T^{\rm out}_{\omega,e} e^{  i k_{ep} x}\right)~; \\[.6\baselineskip]
\theta^{\rm out}_{\omega,2}
&\stackrel{x\to-\infty}{\longrightarrow}
 e^{-i\omega t}\left(e^{  - ik_{wf} x} +
  {\tilde R}^{\rm out}_\omega e^{ i k_{up} x} +
  {\tilde R}^{\rm out}_{\omega,e} e^{  i k_{ep} x}\right)~, \nonumber\\
\theta^{\rm out}_{\omega,2}
&\stackrel{x\to+\infty}{\longrightarrow}
  e^{-i\omega t}
 \left({\tilde T}^{\rm out}_\omega e^{  - i k_{wp} x}\right)~; \\[.6\baselineskip]
\theta^{\rm out}_{\omega,3}
&\stackrel{x\to-\infty}{\longrightarrow}
  e^{i\omega t}\left(e^{  ik_{ef} x} +
 {\bar R}^{\rm out}_\omega e^{  - i k_{up} x} +
 {\bar R}^{\rm out}_{\omega,e} e^{  - i k_{ep} x}\right)~, \nonumber\\
\theta^{\rm out}_{\omega,3}
&\stackrel{x\to+\infty}{\longrightarrow}
 e^{i\omega t}
 \left({\bar T}^{\rm out}_\omega e^{ i k_{wp} x}\right)~.
\end{align}
\end{itemize}
The pictorial representations of these elementary scattering processes
can be seen in Fig.~\ref{Fig:out-basis}. They correspond to the
elementary scattering processes of Fig.~\ref{Fig:in-basis} running
backwards in time (note that for this identification we have to exchange labels 1 and 2).

For the case $\omega<\omega_c$, it is not possible to express the positive norm mode $\theta^{\rm
out}_{\omega,1}$ only in terms of the positive norm modes
 $\theta^{\rm in}_{\omega,1},\theta^{\rm
in}_{\omega,2},\theta^{\rm in}_{\omega,3}$. One has to use the
negative norm modes $\theta^{{\rm in}*}_{\omega,1},\theta^{{\rm
in}*}_{\omega,2},\theta^{{\rm in}*}_{\omega,3}$ as well. Therefore, the
``in'' vacuum state defined through the requirement $a^{\rm
in}_{\omega,i}|0_{\rm in}\rangle=0$ contains quasiparticles coming out
at the right asymptotic region. This phenomenon is usually called mode
mixing and has been thoroughly analyzed in~\cite{Macher:2009tw}.

\subsection{``Stationary'' vacuum state}
\begin{figure}[t]
\includegraphics[width=.45\textwidth]{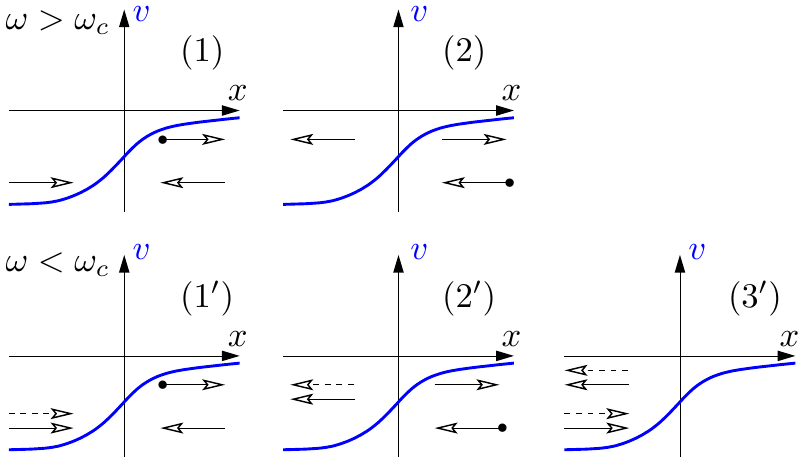}
\bigskip
\caption{\label{Fig:stationary-basis} Preliminary ``stationary'' basis.
\underline{$\omega>\omega_c$:} $(1)$ represents a scattering
process in which there is an outgoing wave to the right;
 $(2)$ represents a  scattering process in which
there is an incoming wave from the right.
\underline{$0<\omega<\omega_c$:} $(1')$ represents a scattering
process in which there is an outgoing wave to the right; $(2')$ represents a  scattering process in which
there is an incoming wave from the right; $(3')$ represents
a  scattering process in which there are neither incoming nor outgoing waves on the right.
These modes are not orthogonal. The ``stationary'' orthonormal basis is obtained by linear combinations of these modes.
  Solid arrows represent ordinary modes while dashed arrows represent extraordinary modes.
}
\end{figure}

Let us finally describe the ``stationary'' set. As a preliminary stage, let
us choose
\begin{itemize}
\item For $\omega>\omega_c$,
\begin{align}
 \theta^\text{st-p}_{\omega,1}=\theta^{\rm out}_{\omega,1}
 \qquad\text{and}\qquad
 \theta^\text{st-p}_{\omega,2}=\theta^{\rm in}_{\omega,2}~;
 \end{align}
 \item For
$0<\omega<\omega_c$,
\begin{align}
 \theta^\text{st-p}_{\omega,1}=\theta^{\rm
out}_{\omega,1}~, \qquad\qquad
\theta^\text{st-p}_{\omega,2}=\theta^{\rm
in}_{\omega,2}~,
\end{align}
 and the additional mode $\theta^\text{st-p}_{\omega,3}$ given by
\begin{align}
\theta^\text{st-p}_{\omega,3}
&\stackrel{x\to-\infty}{\longrightarrow}
 e^{-i\omega t }\left( A_{\omega} e^{  ik_{up} x} + A_{\omega,e} e^{  i k_{ep} x}\right.
 \nonumber\\
&\left.\hspace{12ex}+B_{\omega} e^{  - ik_{wf} x} + B_{\omega,e} e^{  - i k_{ef} x}\right)~, \nonumber\\
\theta^\text{st-p}_{\omega,3}
&\stackrel{x\to+\infty}{\longrightarrow}
 0~,
 \label{eq:st-p3}
\end{align}
if it has positive norm; or by
\begin{align}
\theta^\text{st-p}_{\omega,3}
&\stackrel{x\to-\infty}{\longrightarrow}
 e^{ i\omega t }\left(  A_{\omega} e^{  -ik_{up} x} + A_{\omega,e} e^{-  
i k_{ep} x}\right.
 \nonumber\\
&\left.\hspace{12ex}+B_{\omega} e^{   ik_{wf} x} + B_{\omega,e} e^{   i 
k_{ef} x}\right)~,\nonumber\\
\theta^\text{st-p}_{\omega,3}
&\stackrel{x\to+\infty}{\longrightarrow}
 0~,
\end{align}
if it occurs that the conjugate of the expression in Eq.~(\ref{eq:st-p3}) is the one with
positive norm (this will depend on the specific shape of the velocity profile).  The ratio $A_{\omega}/A_{\omega,e}$ is fixed by
the condition that the scattering lead to no transmission in the right 
asymptotic region.
Then, for a given $A_{\omega}$, $B_{\omega}$ and $B_{\omega,e}$ become fixed.
Finally the value of $A_{\omega}$ itself can be fixed by requiring that 
the mode be normalized.
Fig.~\ref{Fig:stationary-basis} shows a pictorial representation of these 
modes.
\end{itemize}

The previous three modes are clearly independent and normalized but
they are not orthogonal. However, starting from them, it is easy to define
three new modes which are orthonormal. Consider the transformation
\begin{eqnarray}
\left(
\begin{array}{c}
\theta^{\rm st}_{\omega,1} \\
\theta^{\rm st}_{\omega,2} \\
\theta^{\rm st}_{\omega,3}
\end{array}
\right)
=
\left(
\begin{array}{ccc}
M_{11} & M_{12} & M_{13}\\
0 & M_{22} & M_{23}\\
0 & 0 & M_{33}
\end{array}
\right)
\left(
\begin{array}{c}
\theta^\text{st-p}_{\omega,1} \\
\theta^\text{st-p}_{\omega,2} \\
\theta^\text{st-p}_{\omega,3}
\end{array}
\right)~,
\end{eqnarray}
with
\begin{align}
&M_{11}=N^{-1/2}~; \\
&M_{12}=N^{-1/2}(p_{23}p_{13}^*-p_{12}^*)(1-|p_{23}|^2)^{-1}~; \\
&M_{13}=N^{-1/2}(p_{12}^*p_{23}^*-p_{13}^*)(1-|p_{23}|^2)^{-1}~; \\
&M_{22}=(1-|p_{23}|^2)^{-1/2}~; \\
&M_{23}=-p_{23}^*(1-|p_{23}|^2)^{-1/2}~; \\
&M_{33}=1~;
\end{align}
where
\begin{align}
&N=(1-|p_{23}|^2)^{-1}\times \\
&\left(1-|p_{23}|^2 -|p_{12}|^2 -|p_{13}|^2 + p_{12}p_{23}p_{13}^*+ p_{12}^*p_{23}^*p_{13}\right)~; \nonumber \\
&p_{ij}=p_{ji}^* = \langle\theta^\text{st-p}_{\omega,i}|  \theta^\text{st-p}_{\omega,j}\rangle_{\mathcal
 W\text{\!-}\textsc{kg}}~.
\end{align}
It is easy to check that the new modes are
indeed orthonormal. Then the corresponding annihilation operators
transform  with the transposed inverse of the matrix $M$:
\begin{eqnarray}
\left(
\begin{array}{c}
a^{\rm st}_{\omega,1} \\
a^{\rm st}_{\omega,2} \\
a^{\rm st}_{\omega,3}
\end{array}
\right)
=
(M^{-1})^\textsc{t}
\left(
\begin{array}{c}
a^\text{st-p}_{\omega,1} \\
a^\text{st-p}_{\omega,2} \\
a^\text{st-p}_{\omega,3}
\end{array}
\right)~.
\end{eqnarray}
The important point here is that $M$ is an upper
triangular matrix. Therefore, the matrix
$(M^{-1})^\textsc{t}$ appearing in this equation is  lower
triangular. Also, $M^\textsc{t}$, which allows us to write
the annihilation operators $a^\text{st-p}_{\omega,i}$ as linear
combinations of  $a^{\rm st}_{\omega,i}$, is a lower triangular matrix.
This implies that $a^\text{st-p}_{\omega,1},a^\text{st-p}_{\omega,2}$
depend linearly on $a^{\rm st}_{\omega,1},a^{\rm st}_{\omega,2}$ only,
and vice versa, never mixing with   the third mode, which would lead to particle presence
in the right asymptotic region.

Consider the ``stationary'' vacuum state defined by the requirement 
\begin{align}
a^{\rm st}_{\omega,i}|0_\text{st}\rangle=0~.
\end{align}
The important point is that this state does not contain any quasiparticles coming in or out of the right asymptotic region.
To realize that this is indeed so, one has to check that the Bogolubov $\beta$ coefficients defined by the products $\langle\theta^\text{st}_{\omega,i}|  \theta^\text{out *}_{\omega,1}\rangle_{\mathcal W\text{\!-}\textsc{kg}}$ and $\langle\theta^\text{st}_{\omega,i}|  \theta^\text{in *}_{\omega,2}\rangle_{\mathcal W\text{\!-}\textsc{kg}}$ are identically zero ($i=1,2,3$). Given the relation between the $\theta^\text{st}_{\omega,i}$ and the $\theta^\text{st-p}_{\omega,i}$ modes, this follows straightforwardly from the following argument.

By the definition of $\theta^\text{st-p}_{\omega,1}$ and $\theta^\text{st-p}_{\omega,2}$ it is obvious that $\langle\theta^\text{st-p}_{\omega,i}|  \theta^\text{out *}_{\omega,1}\rangle_{\mathcal W\text{\!-}\textsc{kg}}=0$ and $\langle\theta^\text{st-p}_{\omega,i}|  \theta^\text{in *}_{\omega,2}\rangle_{\mathcal W\text{\!-}\textsc{kg}}=0$ for $i=1,2$. In order to show that $\langle\theta^\text{st-p}_{\omega,3}|  \theta^\text{in *}_{\omega,2}\rangle_{\mathcal W\text{\!-}\textsc{kg}}=0$, evaluate the spatial integral that defines this inner product at $t\to-\infty$. Then, it can be seen from the scattering diagrams that the involved modes do not have intersecting support, and therefore their product vanishes.  A similar argument applies for $\langle\theta^\text{st-p}_{\omega,3}|  \theta^\text{out *}_{\omega,1}\rangle_{\mathcal W\text{\!-}\textsc{kg}}$ by evaluating it at $t=+\infty$. (Note that this can be seen even more clearly by using a wave packet basis.) 

Therefore in the ``stationary'' vacuum state there are no quasiparticles coming in or out from the right asymptotic region. In other words, this state is a generalization to this dispersive theory of the Boulware stationary state for black holes and static stars. In standard general relativity, the Boulware state is not regular at the horizon. The dispersive nature of Bogolubov theory, however, makes this state perfectly regular even at the horizon. Consequently, it is a perfectly attainable state in such a dispersive theory.

\subsection{State preparation}
Given a stationary configuration, one could prepare the system to be in
any of the vacuum states described above. The ``in'' vacuum contains
quasiparticles coming out at the right infinity ($x\to+\infty$),  while the
stationary vacuum state does not (it does not contain quasiparticles
coming in from the right infinity either). In standard general relativity,
one can prove that starting from a Minkowskian spacetime at past
infinity and a relativistic field initially in its Minkowski vacuum, if one
dynamically produces a black hole horizon, then the vacuum state at
future infinity is indistinguishable from the Unruh
state~\cite{Wald:1995yp,Racz:1992bp}. In particular, if one takes a Lorentz invariant theory and engineers, by using external means, a quasistatic collapse towards the formation of a black hole, then before the formation of the horizon, one can always make the collapse slow enough so that an initial Minkowskian state first acquires and then 
maintains at each instant of time the Boulware vacuum structure, which is the appropriate vacuum state for a stationary star. However, no matter how slow  the collapse is, the quasistatic approximation breaks down when the horizon forms due to the infinite slow-down of clock rates at the horizon. As a consequence, the Boulware state is unavoidably modified and becomes, after some transient phase, the Unruh state with its associated Hawking emission.
Note that this is also necessary for consistency: the renormalized stress-energy tensor in the Boulware state
is divergent at the horizon and the state itself is not well defined there, because of the infinite blue-shift of the corresponding modes caused by the presence of the (non-dispersive) horizon, as pointed out above. 
The Unruh state, on the contrary, is perfectly regular at the (future)
horizon (we don't mention here states, such as the Hartle-Hawking state,
which are not vacuum states in the past). Thus, within standard general
relativity a field in a black hole spacetime can only be in the Unruh
vacuum state (obviously, modulus the presence of any finite amount of
particles) and never in the Boulware vacuum state.

In dispersive theories, the previous arguments are no longer valid and,
depending on the specific preparation of the configuration---for
instance the specific dynamical way in which one sets up the subsonic to
supersonic transition in the Bose-Einstein condensate flow---one could in principle end up having different
final states for the final stationary flow. Indeed, in a Bose-Einstein condensate there is no infinite delay of the clocks and so it should be possible to maintain the Boulware vacuum structure even after the horizon has been formed. In principle, one could for example end up in the stationary vacuum state described above, which is a
regular extrapolation of the Boulware vacuum  to the dispersive theory. Therefore, although in general the formation of a horizon in a Bose-Einstein condensate leads to a Hawking-like radiation associated to the ``in'' vacuum (as shown numerically in~\cite{Carusotto:2008ep} and studied in detail in~\cite{Macher:2009nz}), it remains to be seen which the precise sufficient conditions are in order to recover the presence of such Hawking radiation in systems with superluminal dispersion relations. This could for instance be relevant when analysing the effect of high-frequency superluminal dispersion in proper black hole configurations and the influence of the black hole's internal region.

This issue has recently created some controversy. In
Ref.~\cite{Barcelo:2008qe} the present authors argued that, under the
assumption that a quasi-static condition for the creation of a black hole
applies,  one does not need to take into account all the modes of the
system to calculate the quasiparticle content at right infinity, but only
those that can be traced as rays escaping from the black hole configuration just before the actual horizon was formed. Quasiparticle production due to the formation of a black hole in
the lab would then appear as just a transient regime and disappears in time: the
system settles down to the stationary state described above.
 However,
other authors claim that the relevant state for the final configuration is
the ``in'' vacuum state (see for example~\cite{Macher:2009tw}). This
vacuum state produces an ever-lasting stationary stream of quasiparticles travelling
towards the right asymptotic region, mimicking in this way a Hawking flux
(the precise spectrum acquires some deviations with respect to a perfect
black body but, in normal situations, these deviations are relegated to the
high-frequency tail of the spectrum). If the ``in'' vacuum state were the
only one available for the configuration, one could conclude that it is not
possible to create a stable sub-to-supersonic transition in the lab: one
would have to maintain it there by external means or it would eventually
dissolve due to backreaction. This is what happens in standard general
relativity:  the existence of stationary black holes is not semiclassically
consistent, they have to evaporate.

 In the Bose-Einstein-condensate dispersive theory, the
availability of the ``stationary'' vacuum implies that,  in principle, it
should be possible to produce a semiclassically-stable analogue of a
stationary black hole in the lab. Which  is the precise vacuum state
selected by the dynamical formation of the black hole horizon configuration
under discussion remains to be calculated.

\section{Conclusions and comments}

When establishing a gravitational analogy in condensed matter systems, a
low-energy Klein-Gordon dynamics emerges for the perturbations around
a background configuration. In this paper, we have addressed the issue of
the appropriate formulation for dealing with the quantum
dynamics of such fluctuations: the one obtained by directly dealing with
the quantum creation and annihilation operators for the excited
quasiparticles (the Bogolubov formalism), or the relativistic approach consisting in obtaining
a generalized Klein-Gordon equation and dealing with it in the
relativistic quantum-field-theoretic way. This question applies in particular
to the quasiparticle Hawking radiation when the background configuration supports an acoustic black hole, either stationary or
externally generated. As we have seen, both methods are entirely
equivalent, leading to the same quasiparticle concept, and hence to the
same description.

It should be stressed that these results are actually more general than
presented here. Indeed, if instead of a Bose-Einstein condensate we had considered
an arbitrary barotropic, inviscid and irrotational fluid described by an
arbitrary enthalpy function $h(n)$, we would have drawn the same
conclusions. Furthermore, for regimes in which the quantum potential is
not significant (long wavelengths), the fluid is described in pure
hydrodynamical terms and a proper Klein-Gordon equation is recovered.
Therefore, we can conclude that the relativistic analogy with a
hydrodynamic fluid is actually more than just an analogy: there is a complete
equivalence leading to a hydrodynamical description of relativistic
massless scalar fields  and vice versa, even at the quantum level and in
strong gravity regimes such as black holes.

Once the inner product is defined one can find a complete and
orthonormal set of positive norm modes, and an associated set of negative-norm ones. One can then
define quasiparticle creation and annihilation operators and make a
choice of vacuum state. This procedure, as is well known, can be carried out in
many equally valid ways, related through Bogolubov transformations. To
illustrate the procedure, we considered a simple configuration of special interest with regard to the possible experimental detection of analogue Hawking radiation: a one-dimensional configuration simulating the presence of a black hole horizon for acoustic perturbations or, in other words, a flow with a subsonic-to-supersonic transition.

In the last section, we have shown that there are several sets
of modes and vacuum states with a particularly simple interpretation: the
``in'' modes and the ``out'' modes (with their associated ``in'' vacuum
and ``out'' vacuum states), already present in previous analyses in the
literature~\cite{Macher:2009tw}, and the ``stationary'' modes and
``stationary'' vacuum state, presented here for the first time. Contrarily
to standard general relativity, the dispersive character of
the Bogolubov theory allows the existence of a regular vacuum state
which does not contain any incoming nor outgoing quasiparticles in the
external asymptotic region, namely, the ``stationary'' state. This state is
a regular generalization of the notion of Boulware state in black hole
physics. Since it is regular, this state can in principle be attained by
adequately preparing the system. In general, the formation of a horizon in a Bose-Einstein condensate should lead to a Hawking-like radiation associated to the ``in'' vacuum~\cite{Carusotto:2008ep,Macher:2009nz}. However, in the light of the present discussion, it should at least theoretically be possible to
produce a black hole analogue in a Bose-Einstein condensate without
causing the emission of a stationary Hawking flux. Whether a black hole
analogue in a Bose-Einstein condensate radiates or not might depend on the
specific path followed in setting up the configuration. This could be an important
issue for experimentalists trying to reproduce Hawking radiation in a lab. From a relativistic point of view, it could also be of considerable importance when analysing the effect of modified dispersion at high energies in gravitational black hole configurations in which the characteristics of the internal region are uncertain. 
We will return to these issues in future studies.

\acknowledgements

Financial support was provided by the Spanish MICINN through the
projects FIS2008-06078-C03-01, FIS2008-06078-C03-03 and
Consolider-Ingenio 2010 Program CPAN (CSD2007-00042)
and by the Junta de Andaluc\'{\i}a through the projects FQM2288 and FQM219. The authors want to thank S.~Finazzi, S.~Liberati, G.A.~Mena Marug\'an and R.~Parentani for illuminating 
discussions.


\end{document}